\documentclass[12pt]{article}
\usepackage{amssymb}
\def\lddots{\mathinner{\mkern1mu\raise1pt\hbox{.}\mkern2mu
\raise4pt\hbox{.}\mkern2mu\raise7pt\vbox{\kern7pt\hbox{.}}\mkern1mu}}
\makeatletter
\def\numberbysection{\@addtoreset{equation}{section}
\def\theequation{\thesection.\arabic{equation}}}
\makeatother

\usepackage{epic}
\usepackage{eepic}
\usepackage{graphicx}
\usepackage{latexsym}
\usepackage{amssymb}
\numberbysection

\newcommand{\be}{\begin{eqnarray}}
\newcommand{\ee}{\end{eqnarray}}
\newcommand{\non}{\nonumber}

\newcommand{\tr}{\mathop{\rm tr}\nolimits}

\begin{document}

\begin{titlepage}
\strut\hfill
%
\begin{center}
{\bf {\Large Generic boundary scattering in the open XXZ chain}}
\\[0.5in]

\vskip 0.2cm

{\large Anastasia Doikou\footnote{e-mail: doikou@bo.infn.it}}

\vspace{10mm}

{\small  University of Bologna, Physics Department, INFN Section\\
Via Irnerio 46, 40126 Bologna, Italy}

\end{center}
%

\vfill

\begin{abstract}

The open critical XXZ spin chain with a general right boundary and a
trivial diagonal left boundary is considered. Within this framework
we propose a simple computation of the exact generic boundary
$S$-matrix (with diagonal and non-diagonal entries), starting from
the `bare' Bethe ansatz equations. Our results as anticipated
coincide with the ones obtained by Ghoshal and Zamolodchikov, after
assuming suitable identifications of the bulk and boundary
parameters.

\end{abstract}

\vfill

\baselineskip=16pt

\end{titlepage}

\section{Introduction}

The derivation of exact bulk and boundary $S$-matrices in the
context of 2-dimensional integrable models is a fundamental
problem that has attracted considerable attention during the last
decades (see e.g. \cite{ZZ}--\cite{friko}). The main aim of the
present investigation is the computation of the exact {\it
generic} --with diagonal and non-diagonal entries-- boundary
$S$-matrix in the framework of the open XXZ spin chain. More
precisely, we consider the open XXZ spin chain with the left
boundary to be trivial ($K^+ \propto {\mathbb I}$), while the
right boundary is associated to a full $K$-matrix \cite{GZ, DVGR}.
The corresponding Hamiltonian reads as: \be {\cal H} &=& -{1\over
4} \sum_{i=1}^{N-1}\Big (\sigma_{i}^{x}\sigma_{i+1}^{x}
+\sigma_{i}^{y}\sigma_{i+1}^{y}+ \cosh (i\mu)\
\sigma_{i}^{z}\sigma_{i+1}^{z}\Big )  - {N\over 4}\ \cosh (i\mu)
-{\sinh (i\mu) \over 4}\sigma_{N}^{z} \non\\
& +& {\sinh (i\mu) \cosh (i\mu \xi) \over 4 \sinh (i\mu \xi)}
\sigma_{1}^{z} - {\kappa \sinh (i \mu) \over 2 \sinh (i\mu \xi)}\Big
(\cosh(i\mu \theta) \sigma_{1}^{x} +i\sinh(i\mu \theta)\sigma_{1}^y
\Big ) \label{H0} \ee where $\sigma^{x,y,z}$ are the $2 \times 2$
Pauli matrices, and the boundary parameters $\xi,\ \kappa,\ \theta$,
are the free parameters of the generic $K$-matrix and  are
associated to some magnetic field applied at the boundaries of the
chain. Notice that we focus here on the critical regime $|e^{i\mu}|
=1$.

In general quantum spin chains are one dimensional statistical
systems displaying particle-like excitations `holes' which may
scatter among each other \cite{FT} or with the boundary \cite{GMN,
done1} in the presence of non trivial boundary magnetic fields. The
boundary $S$-matrix within this framework describes exactly the
reflection of the the particle-like excitation with the boundary
magnetic field. Note that computations concerning both bulk and
boundary $S$-matrices are always valid in the thermodynamic limit of
the spin chain $N \to \infty$ \cite{FT, korepin, AD, GMN, done1,
donebr}. In integrable field theories on the other hand (see e.g.
\cite{GZ}) the physical boundary $S$-matrix describes the reflection
of a solitonic excitation with the boundary, when the system is
considered on the half line \cite{GZ} (see also \cite{ddvb, caux}
for relevant studies on the sine-Gordon model on the finite
interval). It is crucial to note that the critical XXZ model may be
thought of as the discrete analogue of the sine-Gordon model, see
for instance \cite{izko, deve, doikoub} for a detailed derivation of
this correspondence. It was also shown that the bulk \cite{kire,
donebr} and the diagonal boundary $S$-matrices \cite{fesa, donebr,
done1} of the critical XXZ chain coincide with those of the
sine-Gordon model after considering suitable identifications of both
bulk and boundary parameters. It is therefore naturally anticipated
that the generic boundary $S$-matrices should also coincide. Such a
coincidence is also natural from the algebraic point of view since
both models are ruled by the same algebra defined by the reflection
equation. Recall also that boundary $S$-matrices are solutions of
the reflection equation up to an overall physical factor, which in
the spin chain frame may be exactly computed by means of Bethe
ansatz techniques.

There exist several schemes to derive physical boundary
$S$-matrices, such as the bootstrap method developed in \cite{ZZ,
GZ}, the `non-linear integral equation' (NLIE) technique (see e.g.
\cite{ddv, ddvb, nepoh}), and the `physical' Bethe ansatz
formulation \cite{kire, fesa}. However, the most direct means is
arguably provided by the `bare' Bethe ansatz approach (see for
instance \cite{korepin, AD, GMN, done1, donebr}). In this study
the derivation of the generic boundary $S$-matrix is based on the
`bare' Bethe equations \cite{nepo, chin}, and this is the first
time that a direct computation starting from the microscopic Bethe
ansatz equations is achieved for the generic reflection matrix.

The crucial observation is that in the special case under
consideration the Bethe equations reduce to an elegant and
quite familiar form after assuming suitable boundary
parametrizations parallel to the ones of \cite{GZ}. Specifically,
the Bethe ansatz equations acquire a form similar to the one of
the open XXZ spin chain with two diagonal boundaries (see also
\cite{bajn} for relevant comments). We are hence able to compute
the reflection amplitudes in a simple fashion analogous to the
diagonal case, and this is actually one of the main advantages of
the approach adopted here. It is worth pointing out that in the
case where two generic boundaries are implemented the entailed
Bethe ansatz equations are rather involved \cite{nemu, galleas} without
really offering any additional information as far as the boundary
scattering is concerned. Note that in \cite{galleas} the spectrum in
the most generic case with general free bulk and boundary parameters is obtained.
Given the complexity of the Bethe ansatz type equations in the most general case
it is clear that our choice of simpler boundary
conditions is quite practical rendering the relevant computations
considerably more tractable. Moreover, in the particular case we
assume here there exists a non-local boundary conserved quantity,
at first order, contrary to
the generic case. Comments regarding its role on the spectrum and
Bethe ansatz equations, and on its relevance to $S^z$ are also
presented. Our results as expected coincide with the ones obtained
by Ghoshal-Zamolodchikov for the sine-Gordon model with `free'
boundary conditions \cite{GZ}. Relevant results from the NLIE
point of view were also derived in \cite{nepoh}.

\section{The open XXZ spin chain}

We recall the $R$-matrix associated to the the spin ${1\over 2}$
XXZ model, which is a solution of the Yang-Baxter equation
\cite{korepin, baxter}, \be R_{12}(\lambda_1 -\lambda_2)\
R_{13}(\lambda_1)\ R_{23}(\lambda_2) =  R_{23}(\lambda_2)\
R_{13}(\lambda_1)\ R_{12}(\lambda_1 -\lambda_2). \ee The
$R$-matrix has the following form: \be R(\lambda)= \left(
\begin{array}{cc}
\sinh [\mu(\lambda +{i\over 2} +i\sigma^z)]  &\sinh (i\mu)\  e^{\mu \lambda}\ \sigma^- \\
\sinh (i\mu)\  e^{-\mu \lambda}\ \sigma^+   & \sinh [\mu(\lambda +{i\over 2} -i\sigma^z)]  \\
\end{array} \right) \label{Rmat} \ee it is convenient for our purposes here to
set $\mu ={\pi \over \nu}$.

The transfer matrix of the open spin chain is defined as
\cite{sklyanin} \be && t(\lambda) = \tr_{0} \Big \{ M\
K^{+}(\lambda)\  T(\lambda)\
K^{-}(\lambda)\ \hat T(\lambda)\Big \} \non\\
&& T(\lambda) = R_{0N}(\lambda)\ \ldots R_{01}(\lambda),\
~~~~~~\hat T(\lambda) = R_{10}(\lambda) \ldots R_{N0}(\lambda)
\label{transfer2} \ee where $M = diag(q,\ q^{-1})$, $q=e^{i\mu}$
and $K^{\pm}$ are solutions of the reflection equation
\cite{cherednik}: \be R_{12}(\lambda_1 -\lambda_2)\
K_1(\lambda_1)\ R_{21}(\lambda_1 +\lambda_2)\ K_2(\lambda_2)
=K_2(\lambda_2)\ R_{12}(\lambda_1+\lambda_2)\ K_1(\lambda_1)\
R_{21}(\lambda_1-\lambda_2) \label{re}\ee The general solution
$K(\lambda)$ is a $2\times 2$ matrix with entries \cite{GZ, DVGR}:
\be && K_{11}(\lambda)=\sinh[\mu(-\lambda+i\xi)] e^{\mu \lambda},
\qquad K_{22}(\lambda)=\sinh[\mu(\lambda+i\xi)]e^{-\mu \lambda}
\non\\ && K_{12}(\lambda)=\kappa q^{\theta} \sinh(2\mu\lambda),
\qquad K_{21}(\lambda)=\kappa q^{-\theta} \sinh(2\mu\lambda).
\label{def}\ \ee We shall henceforth consider $K^{+} = {\mathbb
I}$, and  $K^-(\lambda)= K(\lambda)$ defined in (\ref{def}). This
particular choice of boundary conditions is compatible with
certain constraints imposed upon the left and right boundary
parameters \cite{chin, nepo, doikouj, doikous}. We are eventually
left with three free boundary parameters associated only to the
right boundary. As a matter of fact the parameter $\theta$ may be
removed via a simple gauge transformation, as also happens in
\cite{GZ}, where a $K$-matrix with $\theta =0$ is assumed for
simplicity, but without loss of generality. Using the fact that
the quantity $(T\ K^-\ \hat T)$ satisfies the reflection equation
one can show that the transfer matrix (\ref{transfer2}) provides a
family of commuting operators \cite{sklyanin}: \be \Big
[t(\lambda),\ t(\lambda') \Big ]=0. \ee

Evaluating the eigenvalues of the $K$-matrix will be particularly
useful for both adopting an appropriate parametrization for our
Bethe ansatz equations as well as for comparing effectively with
the Ghoshal-Zamolodchikov result. Before writing down the
$K$-matrix eigenvalues it will be useful to introduce some
notation: \be {e^{-i\mu \xi} \over 2\kappa } = i \cosh
[i\mu(\beta^- +\gamma^-)], ~~~~~{e^{i\mu \xi} \over 2\kappa } = i
\cosh (i\mu\zeta), ~~~~p^{\pm} = {1\over 2} (\beta ^- +\gamma^-
\pm \zeta) \label{param} \ee compare also with the parametrization
used e.g. in \cite{nepoh, nepo, chin, doikouj, doikous}. Such a
parametrization is also quite natural from the point of view of
boundary Temperley-Lieb algebras \cite{doma}. The latter formulas
(\ref{param}) lead to the following relations among the boundary
parameters: \be {\cosh (i \mu \xi) \over 2 i \kappa} = \cosh (i\mu
p^+)\ \cosh (i\mu p^-), ~~~~~\cosh^2(i\mu p^+) + \cosh^2(i\mu
p^{-}) = 1 -{1 \over 4 \kappa^2} \label{barecon} \ee (c.f. see
similar parametrization in \cite{GZ}, and section 3).
`Renormalized' physical boundary parameters will be ultimately
identified with the boundary parameters of \cite{GZ}, as will
become transparent in section 3. The parameter $\theta$ is hidden
is the sum $\beta^- + \gamma^-$ (see e.g. \cite{chin, doikouj}),
but in any case we shall consider it henceforth to be zero for
simplicity. After diagonalizing the $K$-matrix we obtain the two
eigenvalues: \be \varepsilon_{1,2}(\lambda) = -2 i \kappa\ \sinh [\mu
(\lambda \pm i p^+)]\ \sinh [\mu  (\lambda \pm i p^-)].
\label{eigenb}\ee It is worth stressing that the Bethe ansatz
equations provide essentially the eigenvalues of the physical
boundary $S$-matrix --with `renormalized' boundary parameters.
Nevertheless, the structure of the boundary $S$-matrix is a priori
known due to the fact that is a generic solution of the reflection
equation (\ref{re}). The objective now is to obtain the overall
physical factor in front of the boundary $S$-matrix; this will be
achieved in the subsequent sections by means of the Bethe ansatz
approach.

\subsection{Bethe ansatz equations}

The spectrum and Bethe ansatz equations for the spin ${1 \over 2}$
XXZ chain with general boundary conditions were derived in
\cite{chin, nepo}, whereas in \cite{doikouj, doikous, annecy} the
spectrum and Bethe ansatz equations obtained for various
representations. We shall focus here on the spin ${1\over 2}$
case, the spectrum in this case is given by:
\begin{eqnarray}
&&\Lambda(\lambda) =
K_{1}^{+}(0|\lambda)K_{1}^{-}(p^{\pm}|\lambda)\  \sinh^{2N}
[\mu(\lambda +i)]\ \prod_{j=1}^{M} {\sinh [\mu(\lambda
+\lambda_{j})]\ \sinh [\mu(\lambda -\lambda_{j}-i)] \over \sinh
[\mu(\lambda +\lambda_{j}+i)]\ \sinh
[\mu(\lambda -\lambda_{j})]} \non\\
&+& K_{4}^{+}(0|\lambda)K_{4}^{-}(p^{\pm}|\lambda)\ \sinh^{2N}
(\mu \lambda)\ \prod_{j=1}^{M} {\sinh [\mu(\lambda
+\lambda_{j}+2i)]\ \sinh [\mu(\lambda - \lambda_{j}+i)] \over
\sinh [\mu(\lambda +\lambda_{j}+i)]\ \sinh [\mu(\lambda
-\lambda_{j})]}  \label{ll}
\end{eqnarray}
where $K^{\pm}_{1,4}$, for the particular choice of boundary
conditions are given by (see also Appendix in \cite{doikouj} for the
explicit definitions of $K^{\pm}_{1, 4}$) \be &&
K_1^-(p^{\pm}|\lambda) = -2 i \kappa\ e^{\mu \lambda}\ \sinh [\mu
(\lambda  -i p^-)]\ \sinh [\mu (\lambda -ip^+)],\non\\
&& K_4^- (p^{\pm}|\lambda) =   -2 i \kappa\ e^{\mu \lambda}\ \sinh
[\mu (\lambda + i p^- + i )]\ \sinh [\mu (\lambda + i p^+ + i )]\
{\sinh (2 \mu \lambda)\over \sinh (i\mu)} \non\\ &&
K_1^+(0|\lambda) = e^{-\mu \lambda} \ {\sinh [2\mu (\lambda +i)]
\over \sinh[\mu (2\lambda +i)]}, ~~~~K_4^+(0|\lambda) = e^{-\mu
\lambda} \ {\sinh (i\mu)  \over \sinh [\mu (2\lambda +i)]}. \ee

The Bethe ansatz equations arise as necessary constraints such
that certain `unwanted' terms appearing in the eigenvalue
expression are vanishing. They guarantee also analyticity of the
spectrum, and are written in the familiar form:
\begin{equation} e^{-1}_{2p^- + 1}(\lambda_i;\mu)\
e^{-1}_{2p^+ + 1}(\lambda_i;\mu)\ g_1(\lambda_i;\mu)\
e^{2N+1}_{1}(\lambda_i;\mu) =-\prod_{j=1}^{M}
e_{2}(\lambda_i-\lambda_j;\mu)\
e_2(\lambda_i+\lambda_j;\mu)\ , \label{BAE}
\end{equation} where we define
\be e_{n}(\lambda;\mu) ={\sinh[\mu (\lambda +{in\over 2})] \over
\sinh[\mu (\lambda -{in\over 2})]}, ~~~~~g_{n}(\lambda;\mu)
={\cosh[\mu (\lambda +{in\over 2})] \over \cosh[\mu (\lambda
-{in\over 2})]}. \label{hh0}\ee  The similarity of the Bethe
ansatz equations (\ref{BAE}) with the Bethe equations of the case
with two non-trivial diagonal boundaries is indeed noticeable (see
e.g. \cite{GMN, done1, sklyanin}). This is a crucial observation,
which will considerably simplify the derivation of the generic
boundary $S$-matrix.

In \cite{chin, doikouj, doikous} the spectrum and Bethe ansatz
equations were derived starting from a particular reference state
(the analogue of `spin up' state), however it was shown in
\cite{refst} that a second reference state exists (`spin down'),
providing another set of Bethe ansatz equations similar to
(\ref{BAE}), but with $p^{\pm} \to -p^{\pm}$, guaranteeing also
the completeness of the spectrum. For a detailed analysis on the
completeness of the spectrum using both reference states we refer
the interested reader e.g. to \cite{nemu, refst, annecy}. It
should be stressed that the existence of the two sets of Bethe
ansatz equations together with the similarity of (\ref{BAE}) with
the Bethe equations in the purely diagonal case are the key
elements in deriving the exact boundary $S$-matrix.

Interestingly in the case we are considering here, where the left
boundary is trivial, there exist a conserved quantity, which is
somehow the analogue of the spin $S^z$ of the diagonal case
\cite{sklyanin}. The integer $M$ appearing in the Bethe ansatz
equations is actually associated to the spectrum of the conserved
quantity (see e.g. \cite{doikouj}). Let us briefly recall the
structure of the boundary conserved quantity (boundary non-local
charge), which is defined as \cite{mene1, dema, doikoub, bako}:
\be Q^{(N)} = q^{-{1\over 2} +\theta} K^{(N)}E^{(N)}+q^{{1\over
2}-\theta} K^{(N)}F^{(N)}- {e^{-i\mu \xi} \over 2 \kappa \sinh
(i\mu) } (K^{(N)})^{2}. \label{Q} \ee $K^{(N)}$, $E^{(N)}$ and
$F^{(N)}$ are $N$ coproducts of the quantum algebra ${\cal
U}_{q}(sl_{2})$ \cite{jimbo} i.e. \be && K^{(N)} =
\bigotimes_{n=1}^{N} K_n, ~~~~X^{(N)} =\sum_{n=1}^{N} K^{-1}
\otimes \ldots K^{-1} \otimes  \underbrace{X}_{\mbox{$n$ position}}
\otimes K \ldots \otimes K,
\non\\ && X \in \{E,\ F \}. \label{copp} \ee For the spin ${1\over 2}$
representation in particular ($K \to q^{{\sigma^z \over 2}},\ E
\to \sigma^+,\ F \to \sigma^-$) we have: \be Q^{(N)} = -{i \over \sinh (i
\mu)} \cosh [i \mu(\beta^- +\gamma^- -2 {\cal S})]. \label{as0}
\ee The one-site $Q^{(N)}$ operator becomes a $2\times 2$ matrix
and the eigenvalues of ${\cal S}$ are $\pm {1\over 2}$. For $N=2$,
${\cal S}$ has $0,\ \pm 1$ as eigenvalues, with $0$ being a doubly
degenerate eigenvalue, and so on. It is thus quite clear that the
operator ${\cal S}$ behaves similarly to $S^z$. Details on the
diagonalization of the operator $Q^{(N)}$ for the spin ${1 \over
2}$ representation can be found in \cite{bako}.

From the asymptotic behaviour of the open transfer matrix
(\ref{transfer2}) and the asymptotics of the spectrum (\ref{ll})
the following fundamental formula emerges (for more details on
this matter we refer the reader to \cite{doikouj}): \be Q_{\varepsilon}^{(N)} =
-{i \over \sinh(i\mu)}\ \cosh [i \mu (\beta^- +\gamma^- + 2M-N)].
\label{as} \ee  the subscript $\varepsilon$ in the latter expression stands for the eigenvalue.
In this case $M$ is not fixed, as opposed to the
generic case \label{chin, nepo}, but is associated to the spectrum
of the operator $Q^{(N)}$. From equation (\ref{as}) the upper
(lower) bounds of $M$ (integer) are identified from the spectrum
of the non-local operator $Q^{(N)}$, indeed: \be M = {N \over 2} -
{\cal S}_{\varepsilon}. \label{M} \ee The similarity of the latter relation with
the one appearing in the case of diagonal boundaries is
noticeable; indeed in the diagonal case ${\cal S}$ is simply
replaced by $S^z$ in (\ref{M}). In fact, the symmetry in this case is effectively $U(1)$
resembling the case of purely diagonal boundaries (see also \cite{nic}).
In general the identification of
the spectrum of the boundary non-local charge $Q^{(N)}$ is an
intriguing problem and particular cases have been analyzed in
\cite{bako, nic}.

Finally, comparing the eigenvalues of (\ref{as0}) for an one
particle state ($N$ odd) --the eigenvalues for ${\cal S}$ are then
$\pm {1\over 2}$-- with (\ref{as}) we conclude that: \be M =
{N-1\over 2}. \ee This corresponds to the state with ${\cal S}$ eigenvalue
${1\over 2}$, the other state with eigenvalue $-{1\over 2}$ --above the equator--
may be obtained via a `duality' transformation on the boundary
parameters, as it happens in the diagonal case \cite{sklyanin};
this will be demonstrated however in the subsequent section.

\section{The boundary $S$-matrix}

As already mentioned in the introduction the physical boundary $S$
matrix describes the reflection of a particle-like excitation
(kink), displayed by the XXZ spin chain, with the boundary. Our
main objective here is the computation of the overall physical
factor in front of the boundary $S$-matrix, which contains
significant information regarding the existence of boundary bound
states.

The physical boundary $S$-matrices are denoted henceforth ${\mathrm
K}^{\pm}$. We define the boundary $S$-matrices ${\mathrm K}^{\pm}$
by the quantization condition \cite{AD, GMN} \be \left( e^{i 2
p(\tilde\lambda) N} {\mathrm K}^{+}\ {\mathrm K}^{-}- 1 \right)
|\tilde \lambda \rangle = 0. \label{quantizationopen} \ee $\tilde
\lambda $ is the rapidity of the `hole' --particle-like excitation,
and $p(\tilde \lambda)$ is the momentum of the hole, which will be
formally defined shortly. Recall that both bulk and boundary
scattering may be validly evaluated in the thermodynamic limit
($N\to \infty$) \cite{FT, GMN, done1, donebr}. In this limit as is
well known the so called string hypothesis is valid \cite{FT}. Based
on this hypothesis (see e.g. \cite{FT}) it was shown that the ground
state for the antiferromagnetic spin chain, which we study here,
consists of a `sea' of real (1-string) Bethe roots, and a particle
like excitation is simply a `hole' in this uniformly distributed sea
of real roots \cite{FT, korepin, AD}. For the Bethe ansatz state
with one hole of rapidity $\tilde \lambda$ in the sea, the counting
function --obtained after taking the log of Bethe ansatz equations--
is \be h(\lambda) &=& {1\over 2\pi} \Bigg\{(2 N+1) q_{1}(\lambda\,;
\mu) + r_{1}(\lambda\,; \mu) - q_{2 p_{+}+1}(\lambda\,; \mu)
- q_{2 p_{-}+1}(\lambda\,; \mu) \non \\
&-& \sum_{j=1}^{M} \left[ q_{2}(\lambda - \lambda_{j}\,; \mu) +
q_{2}(\lambda + \lambda_{j}\,; \mu) \right] \Bigg\}, \label{hh}\ee
where \be q_{n}(\lambda;\mu) = \pi +i \log [e_{n}(\lambda;\mu)],
~~~~~~~r_{n}(\lambda; \mu)= i\log[g_n(\lambda;\mu)]. \ee
Note the striking similarity of the latter formula
(\ref{hh}) with the one obtained in
the case of two diagonal boundaries \cite{GMN, done1}. This is a
key point, as already mentioned, allowing us to proceed with a
simplified derivation of the boundary $S$-matrix. The only
difference with the fully diagonal case is that now both terms
depending on $p^{\pm}$ are assigned to the right boundary,
otherwise we proceed exactly as in the diagonal case (see e.g.
\cite{GMN, done1}).

Since we are considering the thermodynamic limit it is clear that
all the sums in the expressions above turn into integrals, and also
we need a density to describe the corresponding states, see e.g.
\cite{FT, GMN, done1, donebr} for details. The Fourier transform of
the density $\sigma_{s}(\lambda)={1\over N} {d h(\lambda)\over d
\lambda} $ describing the one-hole state is given by \be \hat
\sigma_{s}(\omega) &=& 2 \hat  \epsilon(\omega) + {1\over N} {\hat
a_{2}(\omega\,; \mu)\over 1 + \hat a_{2}(\omega\,; \mu)}
(e^{i\omega \tilde \lambda}+ e^{-i\omega \tilde \lambda}) \non\\
&+& {1\over N}{1 \over 1 + \hat a_{2}(\omega \,; \mu)} \bigg[ \hat
a_{1}(\omega \,; \mu) + \hat a_{2}(\omega \,; \mu) + \hat
b_{1}(\omega \,; \mu) - \hat a_{2p^- +1}(\omega \,; \mu)
- \hat a_{2p^+ + 1}(\omega \,; \mu) \bigg], \non\\
\label{dens} \ee where we define the following Fourier transforms
(recall that $|e^{i \mu}| =1$ and $\mu ={\pi \over \nu}$): \be&&
\hat a_n(\omega; \mu) = {\sinh [(\nu-n){\omega \over 2}] \over
\sinh({\nu\omega \over 2})} ~~~~~0< n < 2\nu \\ && \hat b_n(\omega;
\mu) = - {\sinh ({n \omega \over 2}) \over \sinh({\nu\omega \over
2})} ~~~~~~0 < n < \nu \ee and \be \hat \epsilon(\omega) &=& {\hat
a_{1}(\omega\,; \mu) \over 1 + \hat a_{2}(\omega\,; \mu)} = {1\over
2 \cosh \left( {\omega\over 2} \right)}, \ee $\epsilon(\tilde
\lambda)$ corresponds also to the energy of the particle-like
excitation.

The boundary matrix ${\mathrm K}^-$, of the generic form
(\ref{def}), has two eigenvalues ${\mathrm k}_{1,2}$. The left
boundary matrix is trivial \be {\mathrm K}^+(\tilde \lambda) =
k_0(\tilde \lambda) {\mathbb I}. \ee From the density (\ref{dens})
and the quantization condition (\ref{quantizationopen}) we
explicitly derive the quantities $k_0,\ {\mathrm k}_{1,2}$. Indeed
taking into account that \be \epsilon(\lambda)= {1\over 2 \pi} {d
p(\lambda) \over d \lambda} \label{formal}\ee and comparing
(\ref{dens}) with the quantization condition
(\ref{quantizationopen}) we obtain the first eigenvalue ${\mathrm
k}_1$ (see e.g. \cite{GMN, done1} for more details). It is worth
noting that the momentum is not a conserved quantity anymore,
however the momentum of the particle-like excitation may be
formally defined via (\ref{formal}) (see also \cite{GMN, done1}).

To obtain the second eigenvalue ${\mathrm k}_2$ we apply the
argument of \cite{sklyanin, done1} for the diagonal case i.e.
implement a `duality' transformation on the boundary parameters
such that $p^{\pm} \to - p^{\pm}$. This is actually the equivalent
of deriving the Bethe ansatz equations starting from the second
reference state (`spin down') \cite{refst, annecy}. The explicit
expression for the eigenvalue ${\mathrm k}_1$ is given by: \be
{\mathrm k}_1(\tilde \lambda,\ p^+,\ p^-) = {-2 i \kappa \over
\pi^2} \cosh \Big [{\pi \over \nu-1}\Big (\tilde \lambda - {i\over
2} (\nu -2 p^+) \Big ) \Big ] \cosh \Big [{\pi \over \nu-1}\Big
(\tilde \lambda - {i \over 2} (\nu -2 p^-)\Big )\Big ] k(\tilde
\lambda, p^+\ p^-), \non \ee where we define \be k(\tilde
\lambda,\ p^+,\ p^-) = k_0(\lambda)\ k_1(\tilde \lambda,\ p^+)\
k_1(\tilde \lambda,\ p^-) \ee and \be k_0(\tilde\lambda) &=& \exp
\Bigg\{ 2 \int_{0}^{\infty}{d\omega\over \omega} \sinh \left( 2 i
\omega \tilde\lambda \right) {\sinh \left( {3\omega\over 2}
\right) \sinh \left( (\nu - 2){\omega\over 2} \right) \over \sinh
\left( 2\omega \right)
\sinh \left( (\nu - 1){\omega \over 2} \right)} \Bigg\} \non \\
k_1 (\tilde \lambda,\ x)&=& {\pi (-2i\kappa)^{-{1 \over 2}}  \over \cosh \Big
[{\pi \over \nu-1} \Big(\tilde \lambda - {i \over 2} (\nu -2
x)\Big )\Big ]} \exp \Bigg\{ -2 \int_{0}^{\infty}{d \omega\over
\omega} \sinh \left( 2 i \omega \tilde\lambda \right)  {\sinh
\left( (\nu - 2 x - 1) \omega \right) \over 2 \sinh\left( (\nu -
1) \omega \right) \cosh (\omega)} \Bigg\}. \non\\ \label{alphar}
\ee The latter expressions may be written as infinite products of
$\Gamma$-functions, so the comparison with the results in
\cite{GZ} is easier \be k_{0}(\tilde\lambda) &=&
\prod_{n=0}^{\infty} \Bigg\{ {\Gamma \left( {1\over \nu
-1}(-2i\tilde\lambda + 4n + 3) + 1 \right)
\Gamma \left( {1\over \nu -1}(-2i\tilde\lambda + 4n + 1) \right) \over
\Gamma \left( {1\over \nu -1}(2i\tilde\lambda + 4n + 3) + 1 \right)
\Gamma \left( {1\over \nu -1}(2i\tilde\lambda + 4n + 1) \right)}
\non\\ &\times&
{\Gamma \left( {1\over \nu -1}(2i\tilde\lambda + 4n) + 1 \right)
\Gamma \left( {1\over \nu -1}(2i\tilde\lambda + 4n + 4) \right) \over
\Gamma \left( {1\over \nu -1}(-2i\tilde\lambda + 4n) + 1 \right)
\Gamma \left( {1\over \nu -1}(-2i\tilde\lambda + 4n + 4) \right)}\Bigg\} ,
\label{s0r}
\ee
\be
k_{1}(\tilde \lambda,\ x) &=&
\sqrt {i \over 2 \kappa}\prod_{n=0}^{\infty}\Bigg\{
{\Gamma\left( {1\over \nu -1}(-i\tilde\lambda + 2n
-{1\over 2}(\nu - 2 x )) +{1\over 2} \right)
 \Gamma\left( {1\over \nu -1}(-i\tilde\lambda + 2n
+{1\over 2}(\nu - 2x )) +{1\over 2} \right)
\over
\Gamma\left( {1\over \nu -1}(i\tilde\lambda + 2n+2
-{1\over 2}(\nu - 2x)) +{1\over 2} \right)
\Gamma\left( {1\over \nu -1}(i\tilde\lambda + 2n+2
+{1\over 2}(\nu - 2x)) +{1\over 2} \right)}
\non  \\
&\times&
{\Gamma\left( {1\over \nu -1}(i\tilde\lambda + 2n+1
-{1\over 2}(\nu - 2x)) +{1\over 2} \right)
 \Gamma\left( {1\over \nu -1}(i\tilde\lambda + 2n+1
+{1\over 2}(\nu - 2x)) +{1\over 2} \right)
\over
\Gamma\left( {1\over \nu -1}(-i\tilde\lambda + 2n+1
-{1\over 2}(\nu - 2x)) +{1\over 2} \right)
\Gamma\left( {1\over \nu -1}(-i\tilde\lambda + 2n+1
+{1\over 2}(\nu - 2x)) +{1\over 2} \right)}
\Bigg\}. \non  \\
\label{s1r} \ee After implementing the `duality' transformation on
the boundary parameters $p^{\pm} \to - p^{\pm}$ in Bethe ansatz
equations (\ref{BAE}) we obtain the second eigenvalue ${\mathrm
k}_2$ of the reflection matrix with: \be {{\mathrm k}_2 (\tilde
\lambda,\ p^+,\ p^-)\over {\mathrm k}_1(\tilde \lambda,\ p^+,\
p^-)} = {\cosh \left[ {\pi\over \nu - 1} \left( \tilde \lambda +
{i\over 2}(\nu - 2 p^+) \right) \right] \over \cosh \left[
{\pi\over \nu - 1} \left( \tilde \lambda - {i\over 2}(\nu - 2 p^+)
\right) \right]}\ {\cosh \left[ {\pi\over \nu - 1} \left( \tilde
\lambda + {i\over 2}(\nu - 2 p^-) \right) \right] \over \cosh
\left[ {\pi\over \nu - 1} \left( \tilde \lambda - {i\over 2}(\nu -
2 p^-) \right) \right]}, \label{betar} \ee compare also the latter
relation with (\ref{eigenb}); the appearance of renormalized
parameters becomes  now apparent. Notice that the $p^{\pm}$
dependent term, expressed as a product of two $k_1$ functions
(\ref{s1r}), is `double' compared to the diagonal case. In the
diagonal limit only one of the functions $k_1(\lambda,\ p^{\pm})$
survives, while the other one becomes unit (see also relevant
discussion in \cite{GZ}). The diagonal case corresponds to what is
called `fixed' boundary conditions in \cite{GZ}.

Comparison between our findings and the results of \cite{GZ} for
`free' boundary conditions, gives rise to the following
identifications of bulk and boundary parameters. The
Ghoshal-Zamolodchikov bulk coupling constant\footnote{The coupling constant $\lambda$ is
defined in \cite{GZ} as $\lambda = {8\pi \over \beta^2} -1$, where $\beta$
is the familiar sine-Gordon bulk coupling constant.}  $\lambda$
is related to our coupling constant $\nu$ by $\lambda = {1\over \nu - 1}$;
the spectral parameter $u$ in \cite{GZ} is related to our variable
$\tilde \lambda$ by $u =-i \pi \tilde \lambda$. Note that the
original parameter $\mu = {\pi \over \nu}$ appearing in the $R$-matrix (\ref{Rmat}) is
now renormalized to $\tilde \mu = {\pi \over \nu-1}$. Recall also
that the boundary parameters in \cite{GZ} ($\eta,\ \vartheta)$ and $(k,\
\xi'$) (we denote $\xi'$ the Ghoshal-Zamolodchikov $\xi$-parameter
to distinguish it from the bare parameter of (\ref{def})) satisfy
\cite{GZ}: \be \cos (\eta)\ \cosh(\vartheta) = -{1\over k}\
\cos(\xi'), ~~~~\cos^2(\eta) + \cosh^2(\vartheta) = 1 +{1 \over
k^2}, \ee which are similar to the constraints (\ref{barecon})
among the `bare' boundary parameters. Finally, the following
identifications among the boundary parameters are valid: \be
\vartheta=  {i \pi\left( \nu - 2 p^+ \right)\over 2 (\nu -1)},
~~~~~~\eta =  { \pi \left( \nu - 2 p^- \right)\over 2 (\nu -1)},
~~~~~\xi' =  { \pi \left( \nu - 2 \xi \right)\over 2 (\nu -1)},
~~~~~k = -2i\kappa \label{ren}\ee see also relevant formulas in
\cite{nepoh}. With this we conclude our derivation of the generic
physical boundary $S$-matrix, which naturally coincides with the
general reflection matrix found in \cite{GZ} associated to `free'
boundary conditions.

\section{Discussion}

Let us summarize the main findings of this investigation. Our main
objective was the derivation of the exact generic boundary
$S$-matrix for the open XXZ chain. This was achieved by means of
the `bare' Bethe ansatz approach. More precisely, we considered a
particular case of the open spin chain, with a generic right
boundary and a trivial left one. Then assuming appropriate
boundary parametrizations we were able to write down a simple and
familiar form of the Bethe ansatz equations, similar to the Bethe
equations of the XXZ chain with two diagonal boundaries.

The simple form of the Bethe ansatz equations has been a crucial
point in our analysis. This together with the existence of a
second set of Bethe ansatz equations, obtained via a duality
transformation on the boundary parameters, facilitated the
derivation of the two eigenvalues of the generic physical boundary
$S$-matrix. Moreover, we were able to associate the spectrum of
the boundary conserved quantity $Q^{(N)}$ with the spectrum of the
transfer matrix, and we showed that $Q^{(N)}$ plays essentially a
role analogous to $S^z$ in the fully diagonal case. We identified
the overall physical factor in front of the boundary $S$-matrix of
the form (\ref{def}), but with renormalized boundary parameters
(\ref{ren}), which are basically the parameters used in \cite{GZ}.
Thus the reflection matrix derived in \cite{GZ} was fully
recovered.

Similar results may be deduced in the attractive regime where
breathers (bound states) are also present \cite{donebr}.
In this case the computation of the generic
breather boundary $S$-matrix goes along the same lines as in \cite{donebr}.
The $n^{th}$ breather boundary $S$-matrix is then given by:
\be S_b^{(n)}(\tilde \lambda,\ p^{+},\ p^-) =
S_0^{(n)}(\tilde \lambda)\ S_1^{(n)}(\tilde \lambda,\ p^+)\
S_1^{(n)}(\tilde \lambda,\ p^-) \label{breath} \ee
where explicit formulas for $S^{(n)}_{0}(\tilde \lambda),\
S^{(n)}_{1}(\tilde \lambda,\ x)$ are presented in \cite{donebr}.
Expression (\ref{breath}) is analogous to the $n^{th}$ breather reflection matrix in
the sine-Gordon model obtained in \cite{ghos} for `free' boundary conditions. In
the diagonal limit one of the $S_1^{(n)}$ terms in (\ref{breath}) becomes unit,
exactly as in the solitonic reflection matrix, and the result of \cite{donebr}
is recovered corresponding to the `fixed' boundary conditions in \cite{ghos}.
It is clear that such a computation immediately provides the reflection matrix
for the fundamental particle in sinh-Gordon. We recall that the lightest
breather ($n=1$) in sine-Gordon corresponds essentially to the fundamental particle of
the sinh-Gordon model, provided that the coupling constant $\beta \to i\beta$.

Another problem which may be treated in the same spirit is the derivation
of the generic reflection matrix for the XXZ chain in the non-critical regime.
In this case the boundary $S$-matrix will be expressed in terms of
$\Gamma_q$-functions, see e.g \cite{done1, kyoto} for diagonal boundaries only.
Finally an interesting direction to pursue is the computation of generic
boundary $S$-matrices in the context of higher spin open XXZ
chains. Specifically, the spin-$1$ case \cite{kire, spin1} is of
particular significance given its relation to the super-symmetric
sine-Gordon model \cite{super}. We hope to address these issues in
forthcoming publications.
\\
\\
\noindent{\bf Acknowledgments:} This work was supported by INFN, Bologna section, through grant TO12.

\end{document}